CNR
Fosca Giannotti
Mirco Nanni
Luca Pappalardo
Giulio Rossetti
Salvatore Rinzivillo

UNIVERSITY OF PISA
Paolo Cintia
Daniele Fadda
Pietro Luigi Lopalco
Sara Mazzilli
Dino Pedreschi
Lara Tavoschi

WINDTRE
Pietro Bonato
Francesco Fabbri
Francesco Penone
Marcello Savarese

# MOBILE PHONE DATA ANALYTICS AGAINST THE COVID-19 EPIDEMICS IN ITALY

## Flow diversity and local job markets during the national lockdown

# TABLE OF CONTENTS

# INTRODUCTION

Understanding human mobility patterns is crucial to plan the restart of production and economic activities, which are currently put in “stand-by” to fight the diffusion of the epidemics. A recent analysis shows that, following the national lockdown of March 9th, the mobility fluxes have decreased by 50% or more, everywhere in the country [13]. In this report, To this purpose, we use mobile phone data to compute the movements of people between Italian provinces, and we analyze the incoming, outcoming and internal mobility flows before and during the national lockdown (March 9th, 2020) and after the closure of non-necessary productive and economic activities (March 23th, 2020). The population flow across provinces and municipalities enable for the modeling of a risk index tailored for the mobility of each municipality or province. Such an index would be a useful indicator to drive counter-measures in reaction to a sudden reactivation of the epidemics.

Mobile phone data, even when aggregated to preserve the privacy of individuals, are a useful data source to track the evolution in time of human mobility [8, 9], hence allowing for monitoring the effectiveness of control measures such as physical distancing [4, 5, 6]. In this report, we address the following analytical questions: How does the mobility structure of a territory change? Do incoming and outcoming flows become more predictable during the lockdown, and what are the differences between weekdays and weekends? Can we detect proper local job markets based on human mobility flows, to eventually shape the borders of a local outbreak?

*An interactive version of this report will be available at* http://sobigdata.eu/covid_report

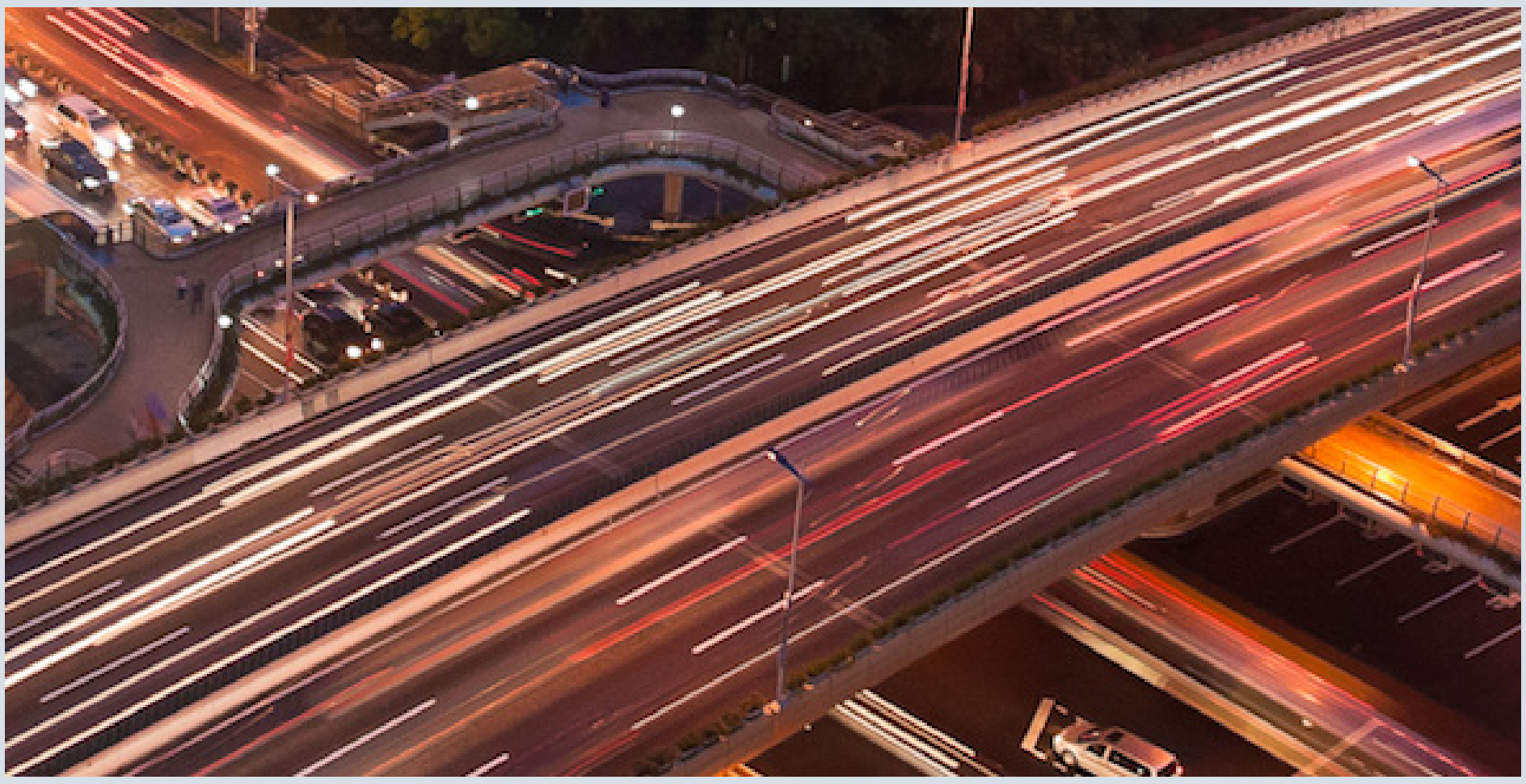

# MOBILE PHONE **DATA**

The raw data used in this report are the result of normal service operations performed by the mobile operator WINDTRE[1]: CDRs (Call Detail Records) and XDRs (eXtended Detail Records). In both cases, the fundamental geographical unit is the "phone cell" defined as the area covered by a single antenna, i.e., the device that captures mobile radio signals and keeps the user connected with the network. Multiple antennas are usually mounted on the same tower, each covering a different direction. The position of the tower (expressed as latitude and longitude) and the direction of the antenna allow inferring the extension of the corresponding phone cell. The position of caller and callee is approximated by the corresponding antenna serving the call, whose extension is relatively small in urban contexts (in the order of 100m x 100m) and much larger in rural areas (in the order of 1km x 1km or more).

Based on this configuration, CDRs describe the location of mobile phone users during call activities and XDRs their location during data transmission for internet access. The information content provided by standard CDR and XDR is the following:

**Call Detail Records (CDR)**

For each phone call, a tuple $<n_o, n_i, t, A_s, A_e, d>$ is recorded, where $n_o$ and $n_i$ are pseudo-anonymous identifiers, respectively of the "caller" and the "callee"; $t$ is a timestamp saying when the call was placed; $A_s$ and $A_e$ are the identifiers of the towers/antennas to which the caller was connected at the start and end of the call; finally, $d$ is the call duration (e.g., in minutes).

**Extended Detail Records (XDR)**

They are similar to CDRs, except that the communication is only between the antenna and the connected mobile phone, and an amount $k$ of kilobytes is downloaded in the process. The format of XDR is, therefore, a tuple $<n, t, A, k>$.

In both CDRs and XDRs, the identity of the users is replaced by artificial identifiers. The correspondence between such identifiers and the real identities of the users is known only to the mobile phone operator, who might use it in case of necessity. This pseudonymization procedure is a first important step (mentioned in Article 6(4) and Article 25(1) of the GDPR, the EU General Data Protection Regulation) to provide anonymity [7, 10, 11] and it will then turn into totally anonymous data for the possible treatment data use. For the analyses in this report, we used aggregated data computed by the mobile operator covering the period February 3rd, 2020 to March 28th, 2020.

[1]WINDTRE is one of the main mobile phone operators in Italy, covering around 32% of the residential "human" mobile market.

# ORIGIN-DESTINATION **MATRICES**

CDRs and XDRs are aggregated into daily municipality-to-municipality origin-destination (OD) matrices: there is an OD matrix per each day, and each element $OD_{A,B}$ of the matrix describes the total number of trips from municipality $A$ to municipality $B$. The presence of two consecutive points of a user in different municipalities indicates a movement, which is counted as a trip if the user stays in the destination municipality for at least one hour, and discarded otherwise. For a better matching with public COVID-19 data, we aggregated the municipality-to-municipality ODs into province-to-province ODs, in which each node represents an Italian province. The trips between municipalities of the same province have been aggregated into a self-loop, which indicates the province's internal mobility. As they are calculated by the operator, we store the daily municipality-to-municipality OD matrices and the daily province-to-province ones into a relational DBMS and access them through calls to a dedicated API.

Figure 1 visualizes the out-flows and in-flows of the province of Padua (region of Veneto, north-east of the country), for February 18th (before the lockdown, on the left) and March 24th (during the lockdown, on the right). The chart shows the flows among provinces with a stroke width proportional to the flows. The out-flows (in-flow) are first linked to the corresponding region and then to the final destination (origin). During the lockdown, we observe a drastic reduction of both the in- and the out-flows (reported on labels in the corresponding circle), as well as a reduction of the number of provinces the flows are coming from or are directed to. The reduction in the number of provenances and destinations is also evident in the other provinces of the country. For example, Figure 2 shows this pattern is even more pronounced for the province of Bari, in the region of Puglia in the south-east of the country.

# BEFORE LOCK DOWN
# **DATA**

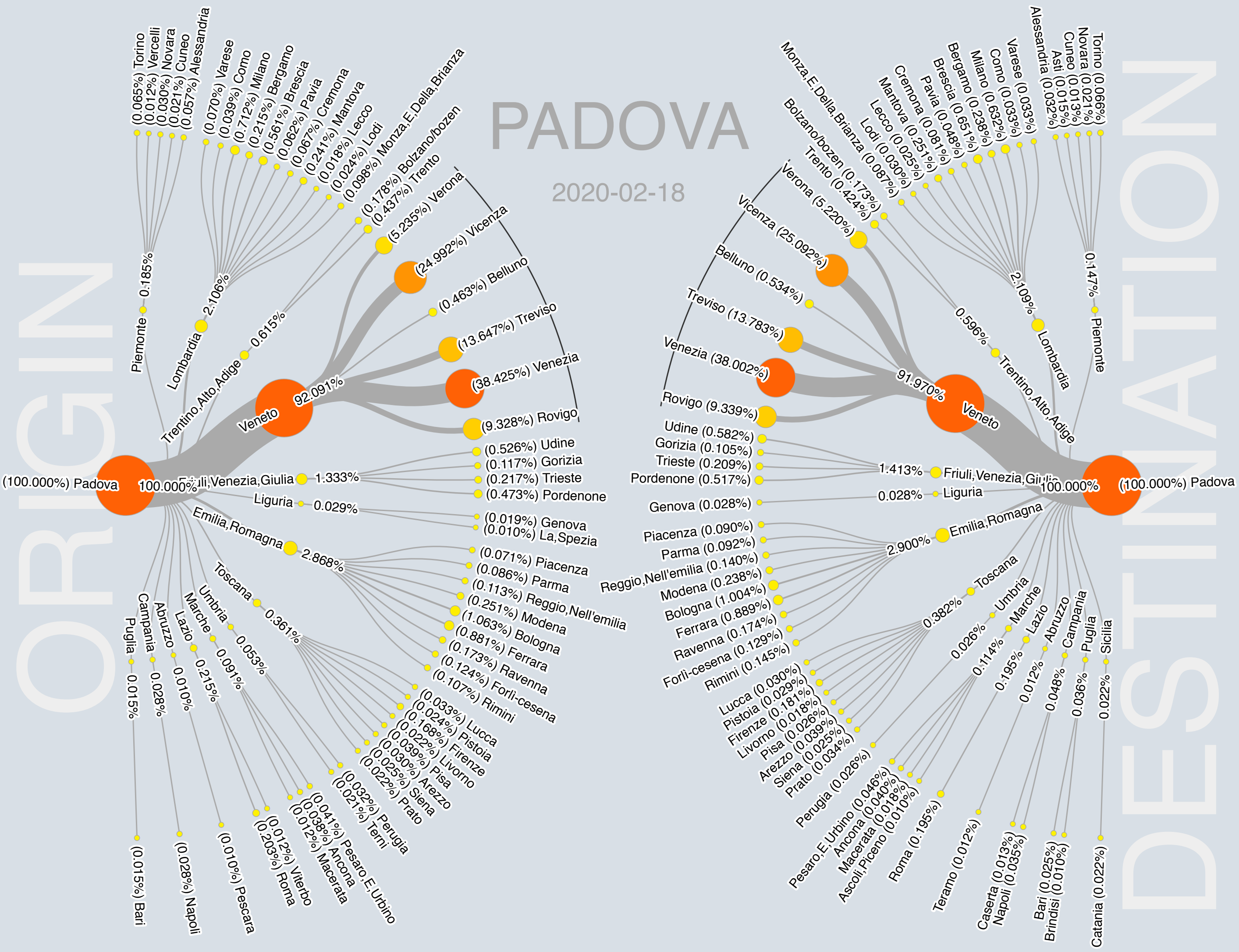

# DURING LOCK DOWN **DATA**

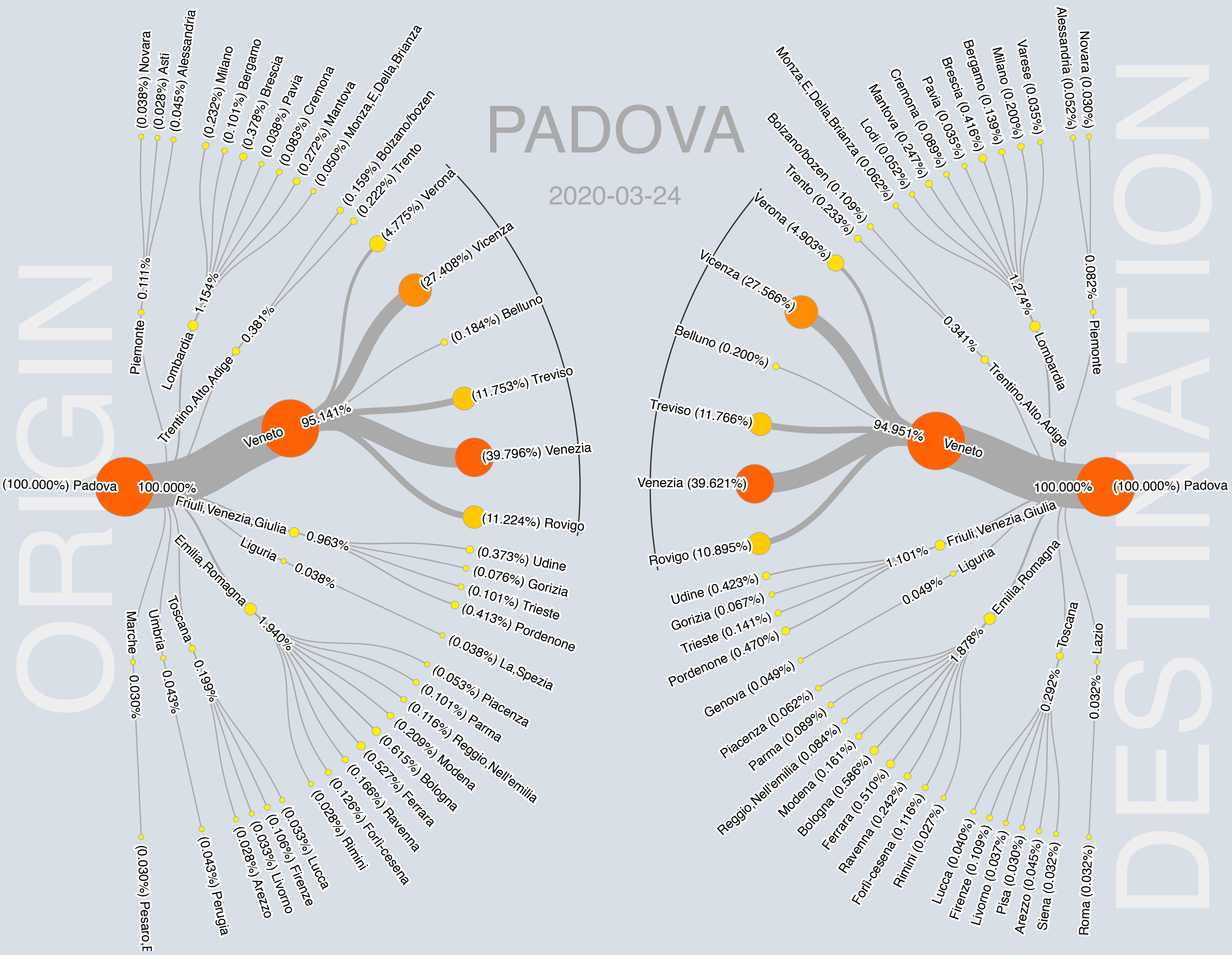

**FIGURE 1 - PROVINCE OF PADUA**

Visualization of the in-flows and the out-flows of the province of Padua (region of Veneto, north-east of the country), on Tuesday February 18th (before the lockdown, on the left) and Tuesday March 24th (during the lockdown, on the right). Note that most of the flows are contained within the Padua's region (Veneto) and neighboring regions, that the number of distinct origins and destinations of flows decrease during the lockdown.

# BEFORE LOCK DOWN **DATA**

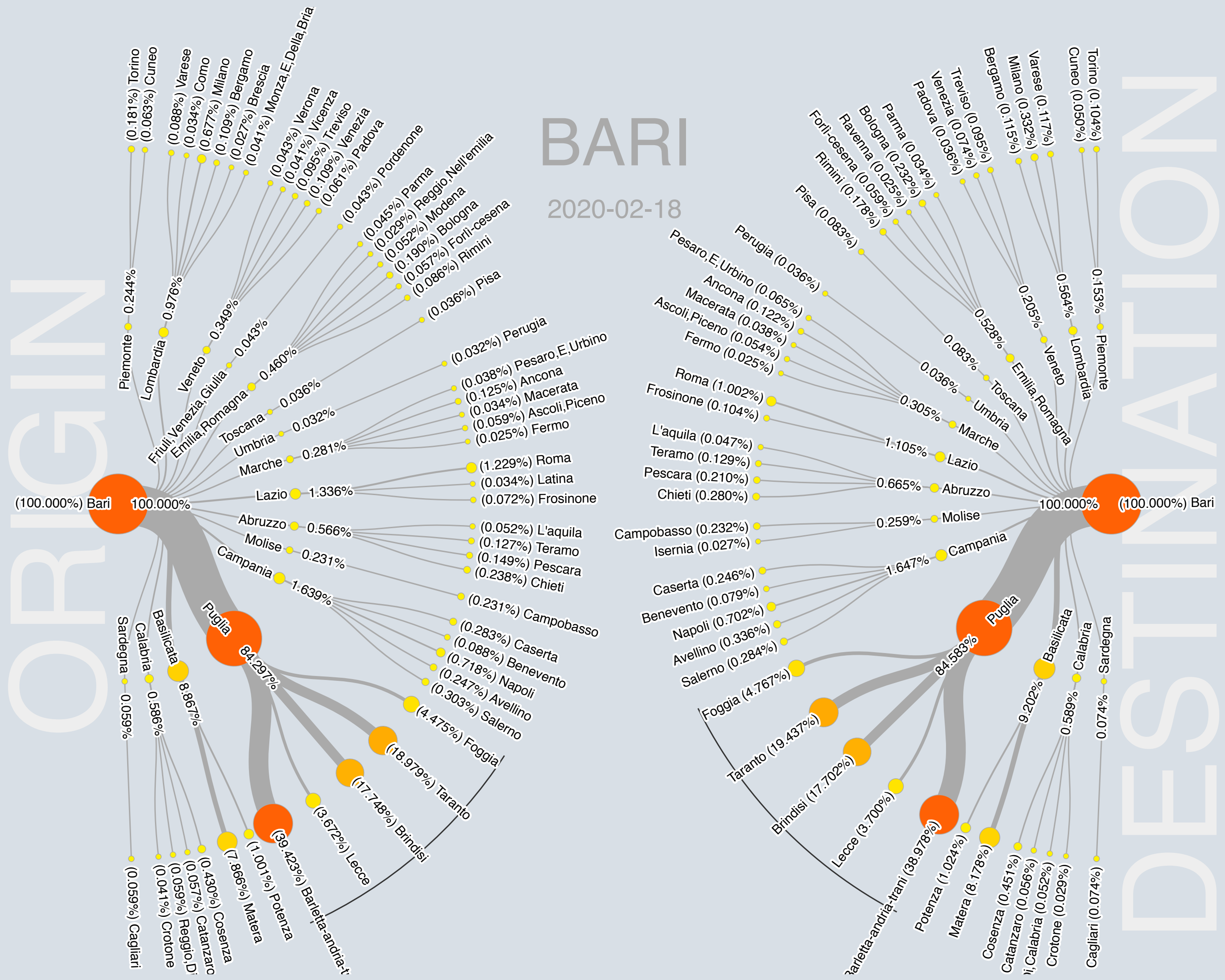

# DURING LOCK DOWN **DATA**

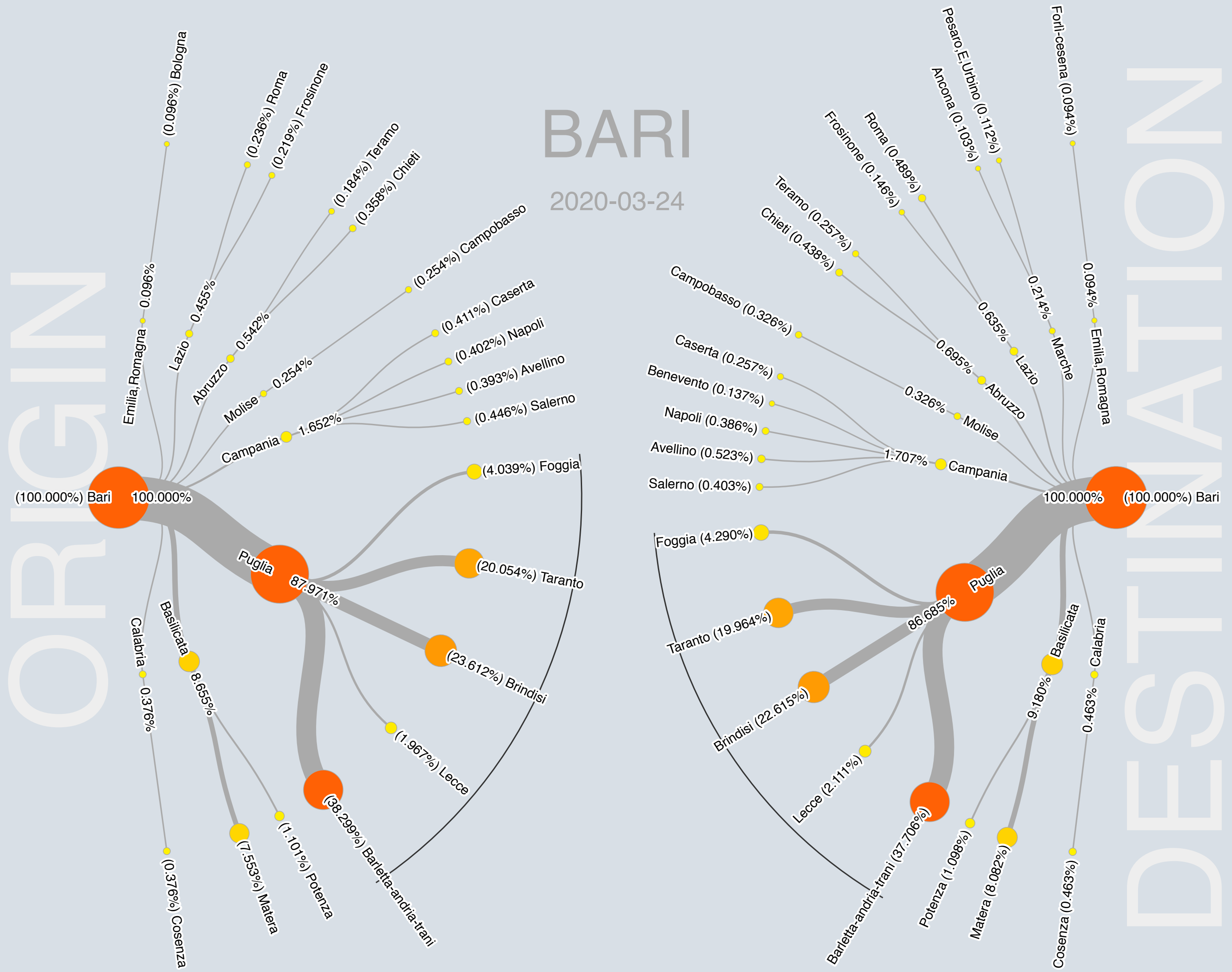

**FIGURE 2 - PROVINCE OF BARI**

Visualization of the in-flows and the out-flows of the province of Bari (in the region of Puglia, south-east of the country), February 18th (before the lockdown, on the left) and March 24th (during the lockdown, on the right). Note the drastic reduction in the number of distinct origins and destinations during the lockdown.

# INCOMING, OUTCOMING AND **INTERNAL MOBILITY FLOWS**

How did mobility flows in Italy change during the lockdown? This aspect is crucial to quantify to what extent the government directives have had the desired effect. We selected four Italian provinces: Bergamo, Padua, Bari, and Catania. For each of these provinces, we computed the evolution day by day of three types of flow:

1. out-flows, indicating the total number of people moving **from** the province to any other province in Italy on that day;
2. in-flows, indicating the total number of people moving **to** the province from any other province in Italy on that day;
3. self-flows (or internal flows), indicating the total number of people moving between municipalities of the same province on that day.

Figure 3 shows the evolution of the normalized in-flows, out-flows, and self-flows of the selected provinces. It is evident how all provinces have a net decrease of the in-, out- and self-flows soon after the first national lockdown (March 9th), and a stabilization of the flows on the new volume after around one week, from March 15th. Therefore, subsequent ordinances, such as closing factories on March 17th, have had a minor impact on the reduction of mobility flows.

**FIGURE 3 - EVOLUTION OF FLOWS**

The two vertical lines indicate the dates of the national lockdown (March 9th) and the closure of non-necessary productive and economic activities (March 23th). We observe a significant decrease in the volume of flows after the national lockdown, while we do not observe a comparable decrease soon after the closure of non-necessary activities.

# NORMALIZED FLOWS WITH RESPECT TO THE MAXIMUM VOLUME OBSERVED

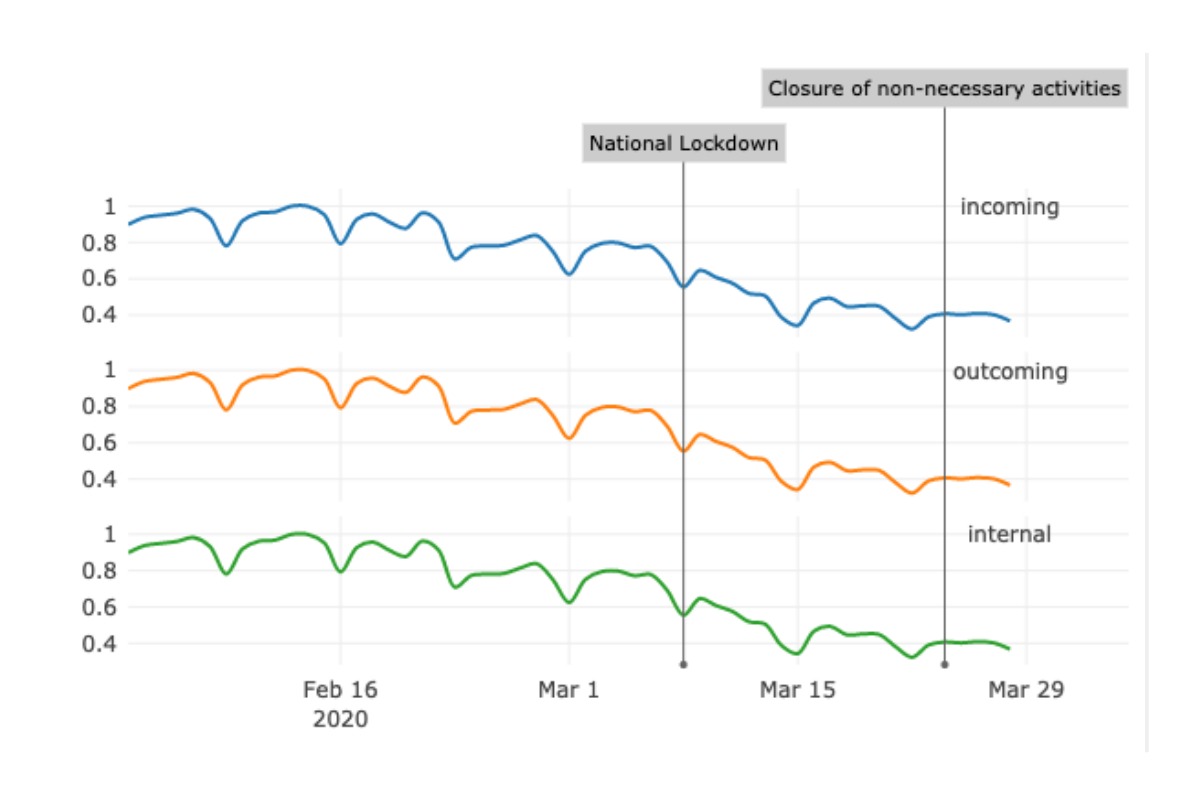

## BERGAMO

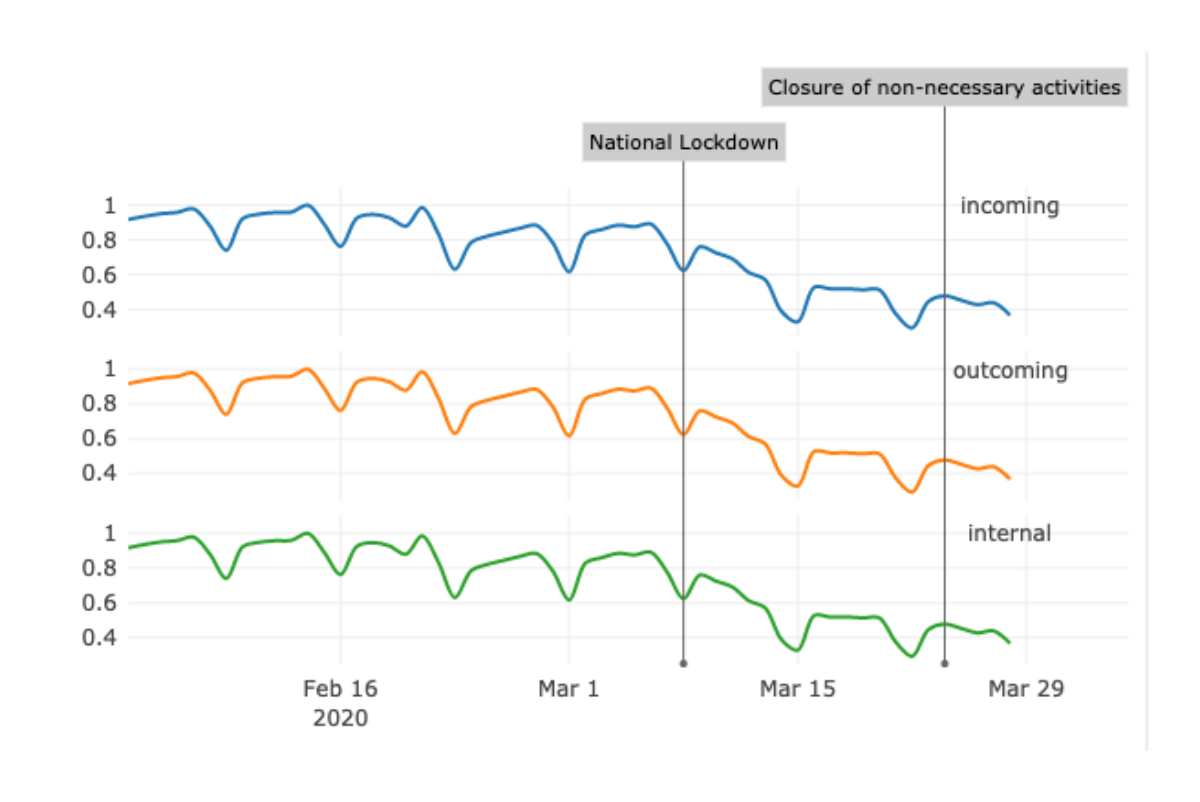

## PADUA

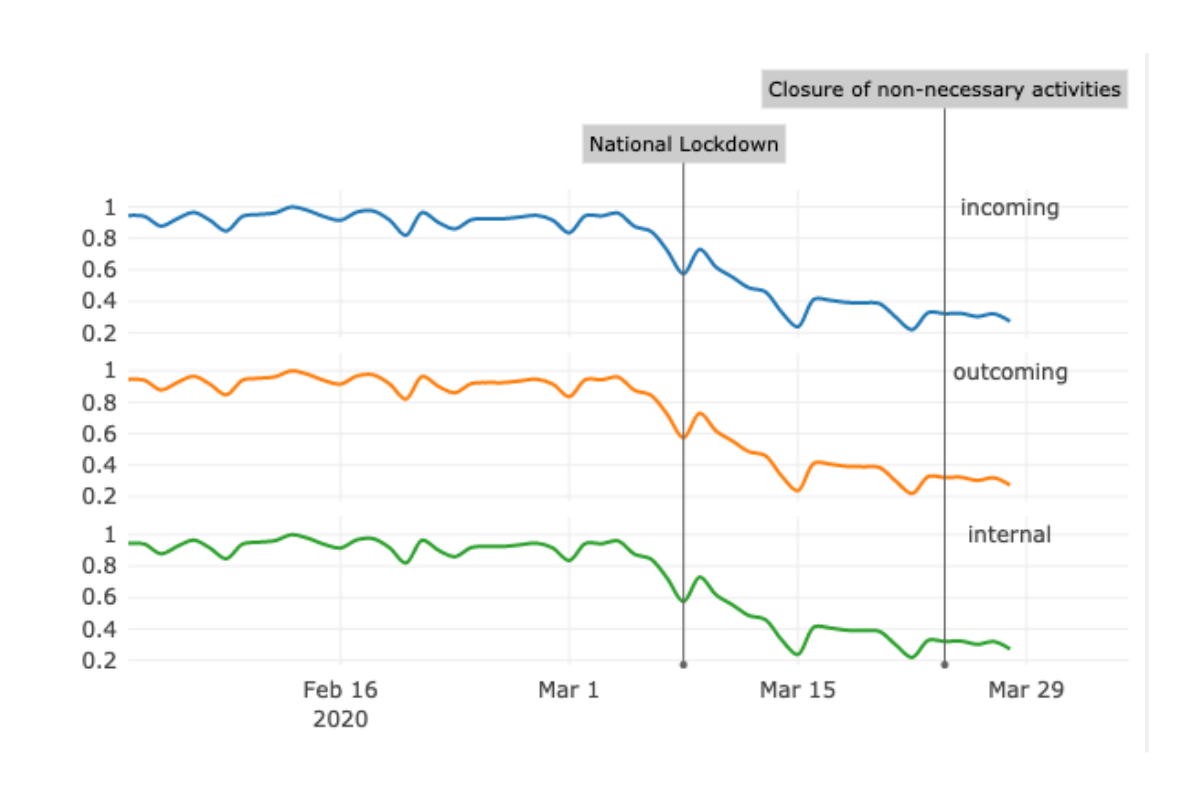

## BARI

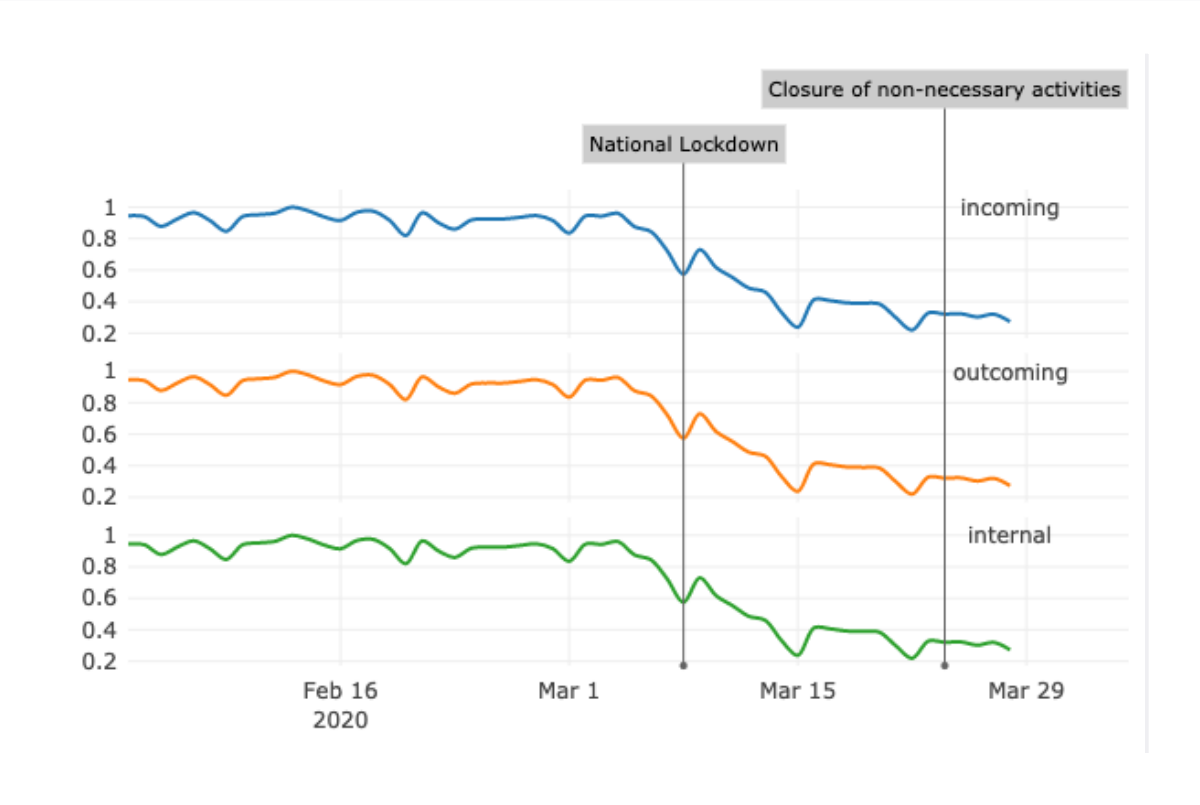

## CATANIA

# FLOW **DIVERSITY**

An important aspect of the mobility of a province is the diversification of the provenience and the destination of people. Specifically, we define the in-flow diversity of a province $A$ as the Shannon entropy of the in-flows to the province [12]:

$$E_{in}(A) = -\frac{\sum_{x \in P_{in}(A)} p(x) \log p(x)}{\log (N)}$$

where $P_{in}$ is the number of provinces with non-null flow to province $A$, $p(x)$ is the probability that the in-flow to province $A$ comes from province $x$, and $log(N)$ is a normalization factor where $N=110$ is the number of Italian provinces. The out-flow diversity of province $A$ is computed similarly as:

$$E_{out}(A) = -\frac{\sum_{x \in P_{out}(A)} p(x) \log p(x)}{\log (N)}$$

where $P_{out}$ is the number of provinces with non-null flow from province $A$, and $p(x)$ is the probability that the out-flow from province $A$ goes to province $x$.

The horizon charts in Figure 4 show the evolution of the in- and out-flow diversity for the four selected provinces, while those in Figure 5 refer to 30 provinces chosen randomly. The vertical axis lines represents time, each rectangle section has a color proportional to the displayed measure (darker color for larger value). The circles on the left have an area proportional to the number of confirmed COVID-19 cases in the corresponding province up to March 24th. We find a progressive reduction of both the in- and out-flow diversity as time goes by, with an acceleration of the process soon after the beginning of the national lockdown (March 9th). Before the lockdown, the in- and out-flow diversities are slightly higher at the weekends than the weekdays. The opposite is true during the lockdown: the in- and out-flow diversities are considerably lower at the weekends than the weekdays. This exciting result suggests that: *(i)* the provenience and the destination of a province's mobility flows during the lockdown are more predictable than before the lockdown; *(ii)* regarding the weekends, the provenience and destination of flows are more diverse before the lockdown than during it.

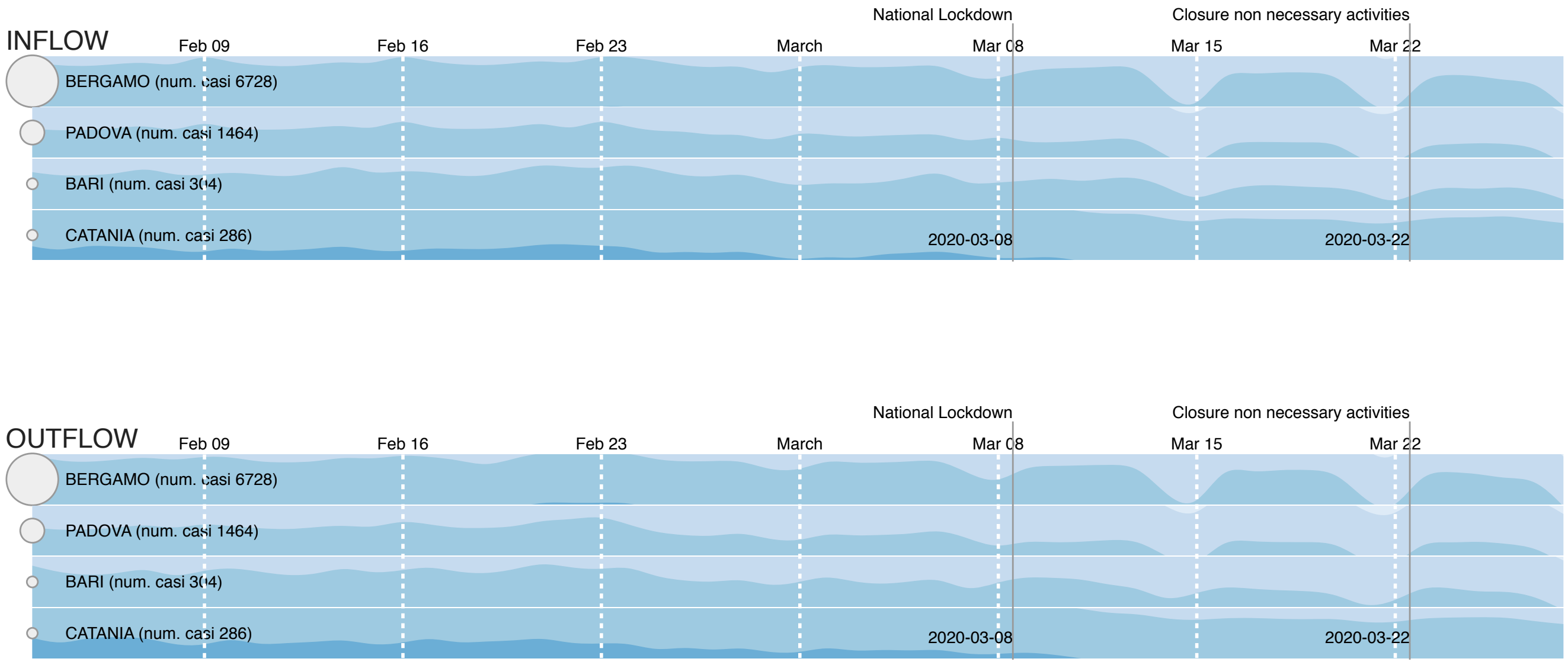

**FIGURE 4 - EVOLUTION OF FLOW DIVERSITY**

Horizon chart that describes the evolution in time of the in- and out-flow diversity of the provinces of Bergamo, Padua, Bari and Catania. The circles on the left have an area proportional to the number of confirmed COVID-19 cases in the corresponding province up to March 24th. Horizon charts compact the area chart by slicing it horizontally, and then shifting the slices to baseline zero. Black solid vertical lines indicate the dates of the national lockdown (March 9th) and the closure of non-necessary productive and economic activities (March 23th). The white dashed vertical lines indicate Sundays. Note that, while the in- and out-flow diversities slightly increase in the weekends before the lockdown, they decrease in the weekends during the lockdown.

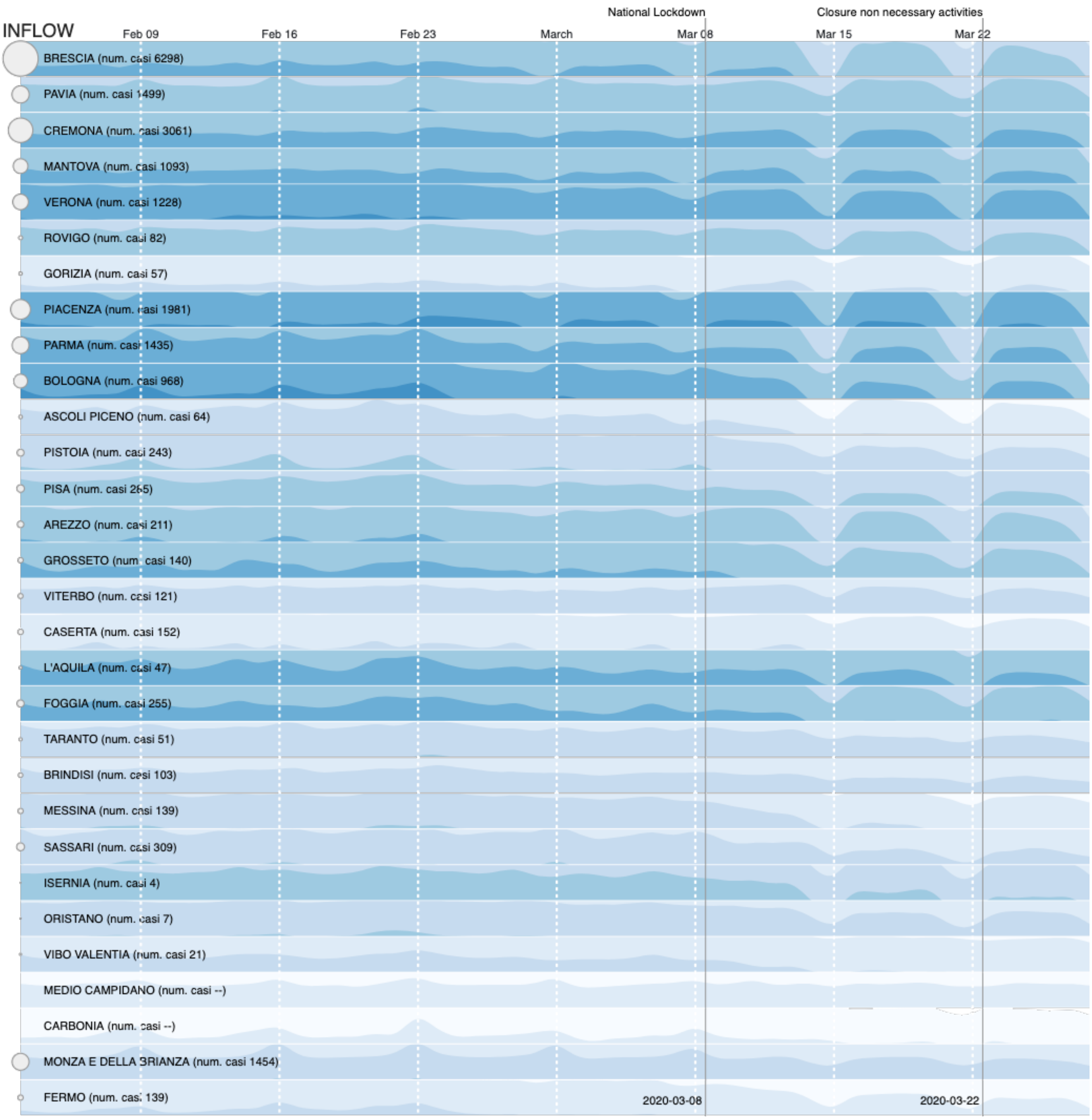

**FIGURE 5 - EVOLUTION OF FLOW DIVERSITY**

Horizon chart that describes the evolution in time of the in-flow diversity of 30 provinces (out of 110) chosen at random. Solid vertical lines indicate the dates of the lockdown (March 9th) and the closure of non-necessary economic activities (March 23th). The dashed vertical lines indicate Sundays. We observe an interesting pattern for weekends: while flow diversity slightly increases with respect to weekdays before the lockdown, it decreases during the lockdown.

# CLUSTERS **OF PROVINCES**

We use the *k-means* clustering algorithm to discover $k$ groups of similar provinces in terms of their evolution of in- and out-flow diversity. To find the best value of $k$, we repeat the algorithm for $k = 2, ..., 20$. For both the in- and out-flow diversities, we find that $k = 5$ minimizes the within-cluster distance. Figure 6 shows the centroids of the five clusters of in-flow diversity.

Although the clusters' trends are similar, they have different typical in-flow diversities. We provide in the Appendix the figure that shows the clusters' centroids regarding the evolution of the out-flow diversity. Figure 7 visualizes the evolution of the in-flow diversity for all the provinces in cluster 4, the one with the highest typical in-flow diversity. We provide in the Appendix the horizon charts regarding the other four clusters.

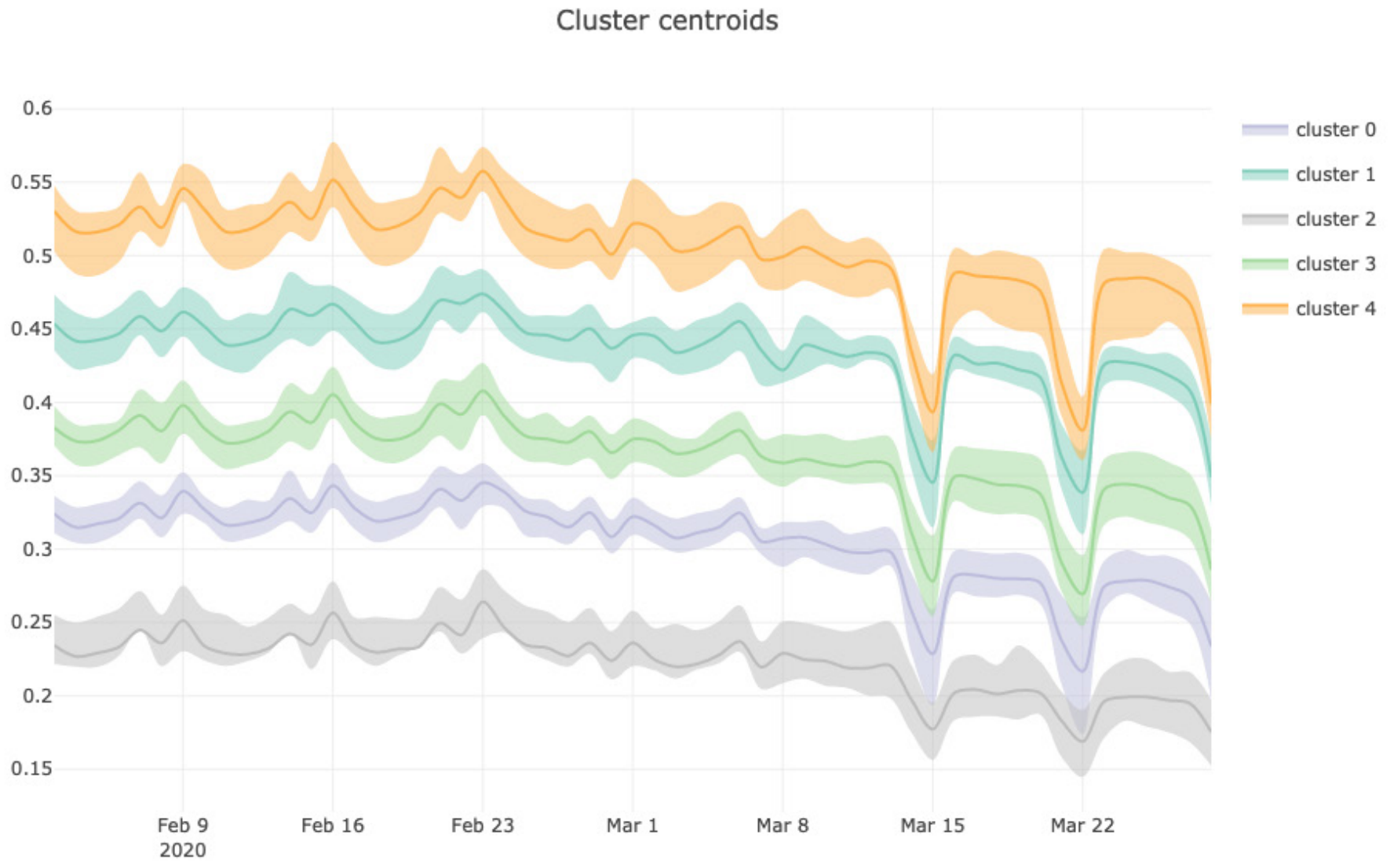

**FIGURE 6 - CLUSTERS OF PROVINCES**

Evolution of the in-flow diversity of the five clusters' centroids. The area around the line indicates the deviation of the provinces from the centroid. Note that, though the clusters have similar trends, they have different typical in-flow diversities.

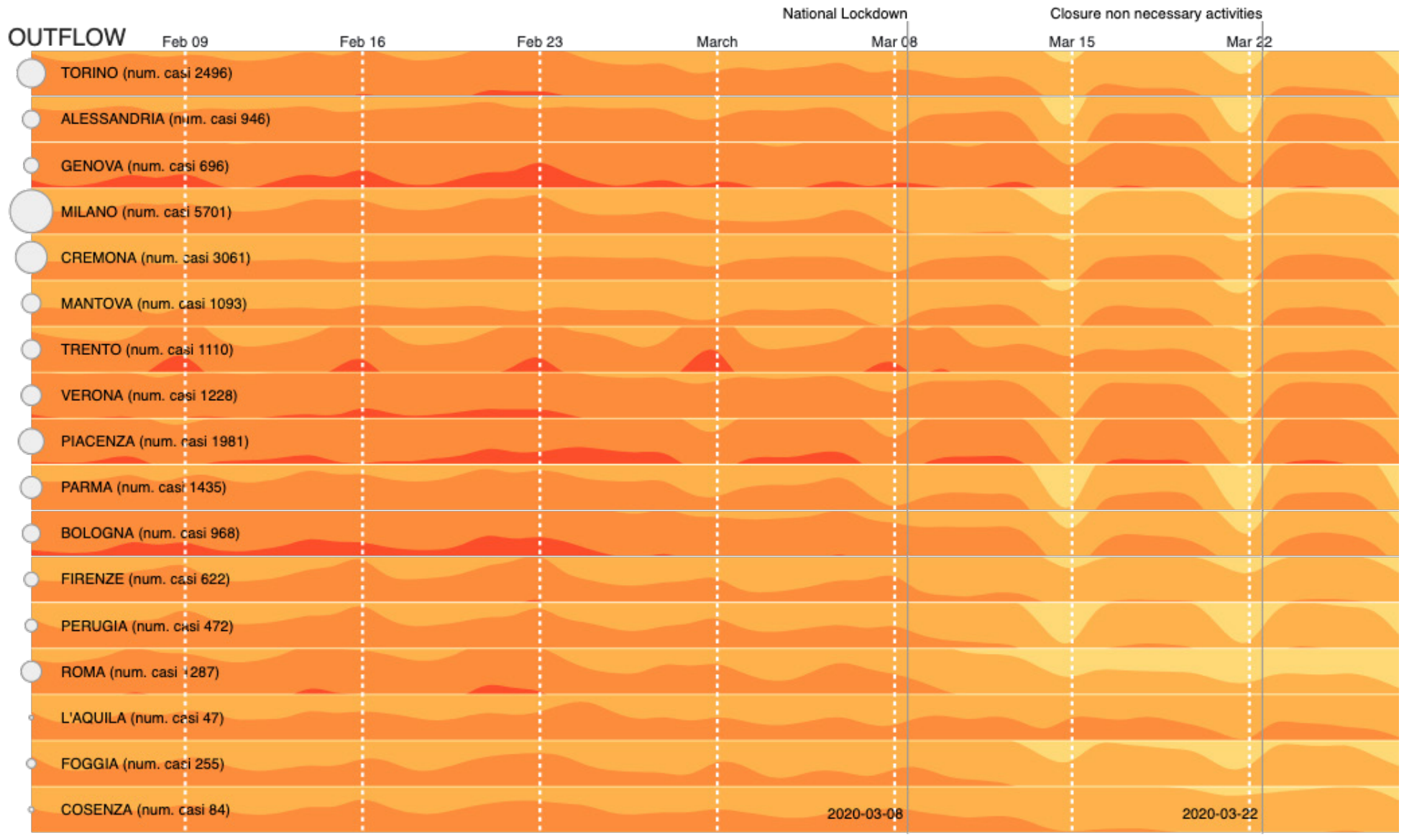

**FIGURE 7 - EVOLUTION OF IN-FLOW DIVERSITY OF CLUSTER 4**

Evolution of the in-flow diversity of provinces in cluster 4. This cluster includes Bergamo and Brescia, territories among the most hit by COVID-19. The circles on the left have an area proportional to the number of confirmed cases in the corresponding province up to March 24th.

# LOCAL JOB **MARKETS**

Economic activities are linked by input-output relationships, with interconnected supply chains that are difficult to isolate. Local Job Markets (LJM) take into account the shifts between the home-work displacements (commuting) that occur between different municipalities. Each LJM is partially isolated from the others, allowing from a more precise control with respect to the administrative classifications of the territory (e.g., municipalities, provinces, regions).

One strategy to detect LJMs is to identify clusters of geographical areas that are homogeneous w.r.t. their mobility flows, i.e., groups of municipalities for which the internal mobility fluxes are denser than the outgoing ones. This can be done by analyzing the municipality-to-municipality OD matrices as weighted directed graphs [1] and using a community detection algorithm [2] to discover a collection of well-bounded mesoscale topologies, e.g., municipality clusters. Note that community detection algorithms can provide different results depending on their definition of what a community is [2]. We use algorithm Infomap [3], which uses the probability flow of random walks on a graph as a proxy for information flows in the real system, and decomposes the network into clusters by compressing a description of the probability flow. The algorithm looks for a cluster partition of the given network that minimizes the expected description length of a random walk.

We applied Infomap to the daily municipality-to-municipality OD matrices. Figure 8 shows the evolution in time of the number of LJMs (communities) in the country. Since the national lockdown (March 9th), there has been a striking increase in the number of communities, indicating that people moved within smaller areas. For example, note that on Monday, March 2nd (before the lockdown), we have around 350 communities, while the number of communities on Monday, March 16th (during the lockdown) is around 550 communities, 200 more. We also find that, during the lockdown, the number of communities increases by around 50% on the weekends (in Figure 8, Sundays are denoted by vertical lines). Also, note that the number of communities increases after the closure of non-necessary productive and economic activities (March 22nd). This may indicate that, while we do not appreciate any significant difference in the volume and diversity of flows after this closure, the structure of mobility flows has changed significantly.

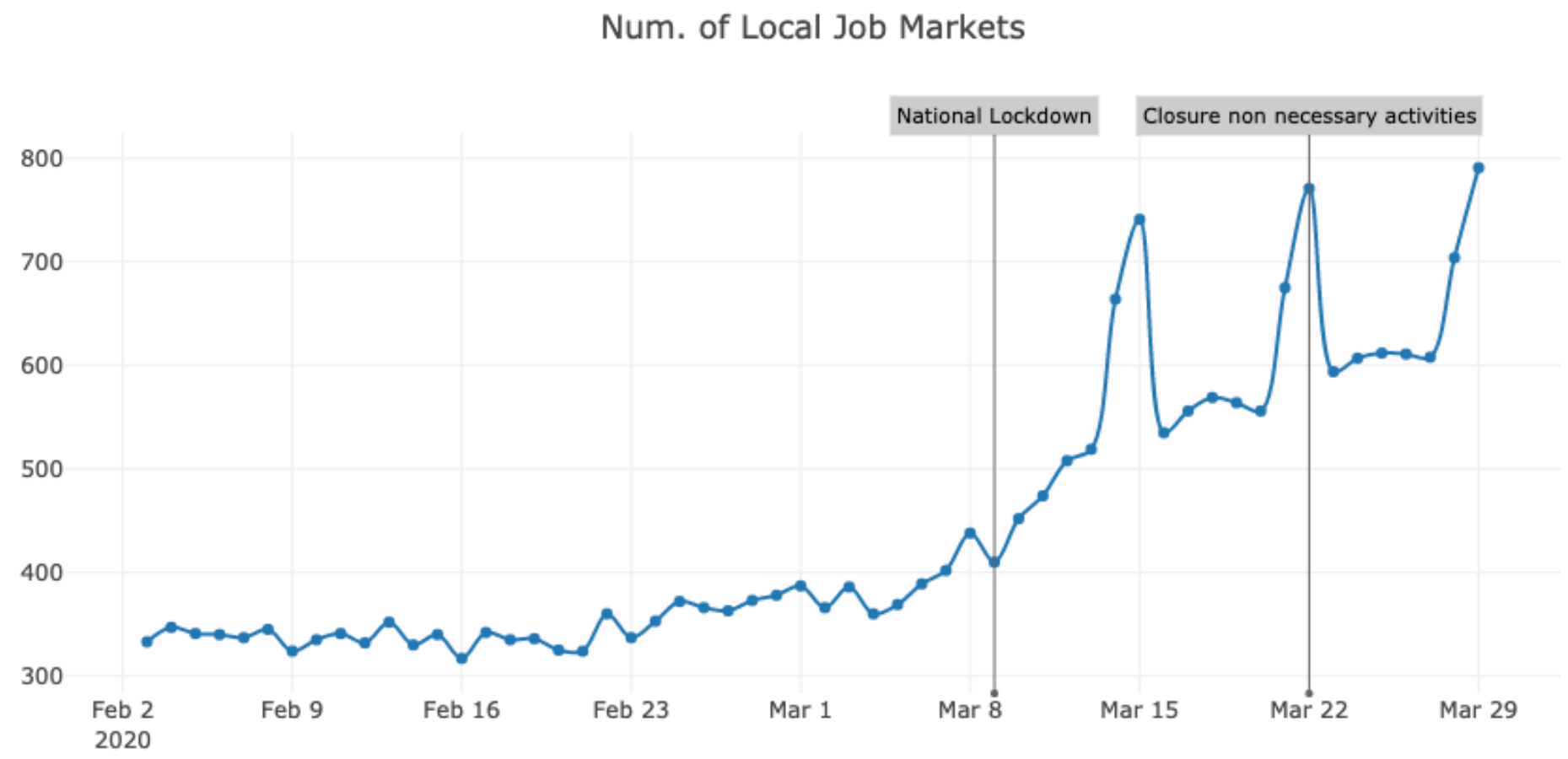

**FIGURE 8 - NUMBER OF LOCAL JOB MARKETS IN TIME**

Evolution in time of the number of Local Job Markets (communities) in Italy, according to the Infomap algorithm. Grey vertical lines indicate Sundays. Note that, after the beginning of the national lockdown (March 9th), there is a striking increase of the LJMs. Moreover, there is a slight increase of the communities after the closure of non-necessary activities (March 22nd).

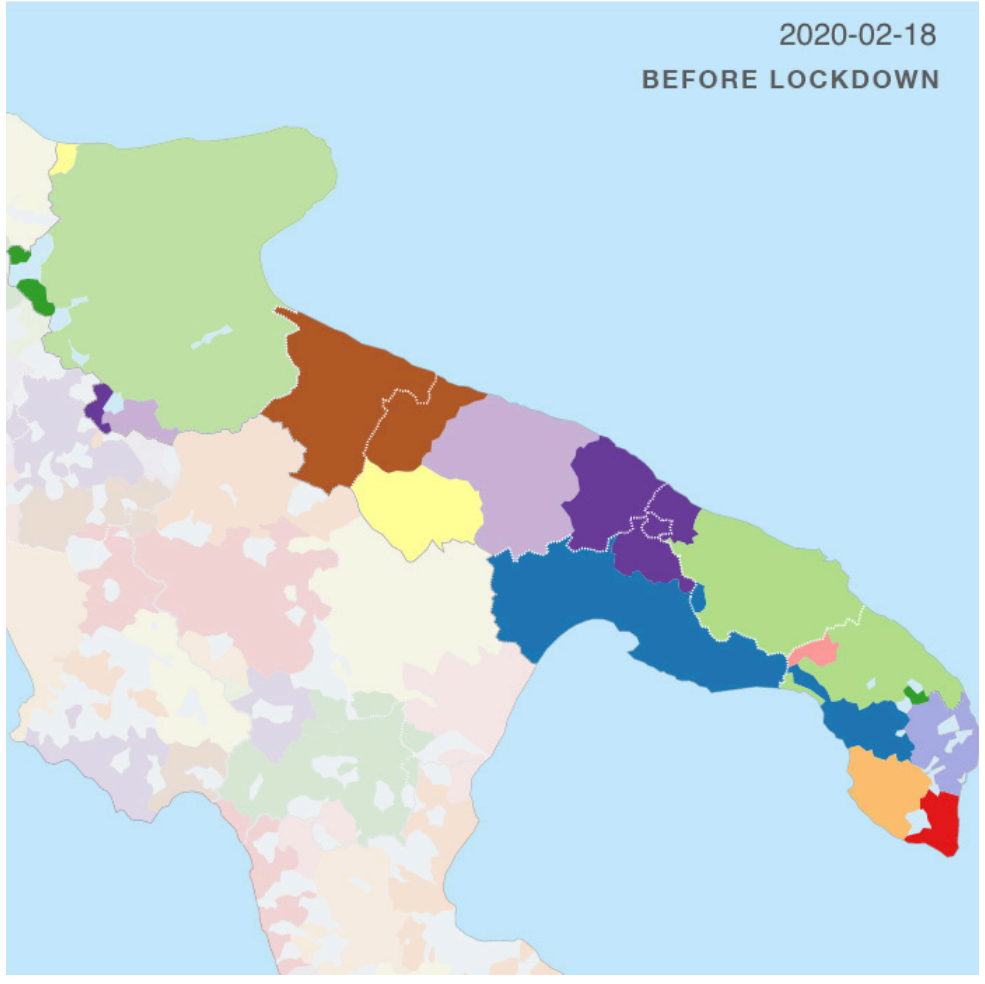

Figure 9 shows the local job markets we found in Puglia (a region on the south-east of the country) before the lockdown (up) and during the lockdown (down). Note the fragmentation of the territory during the lockdown, especially for the easternmost and the westernmost parts of the region.

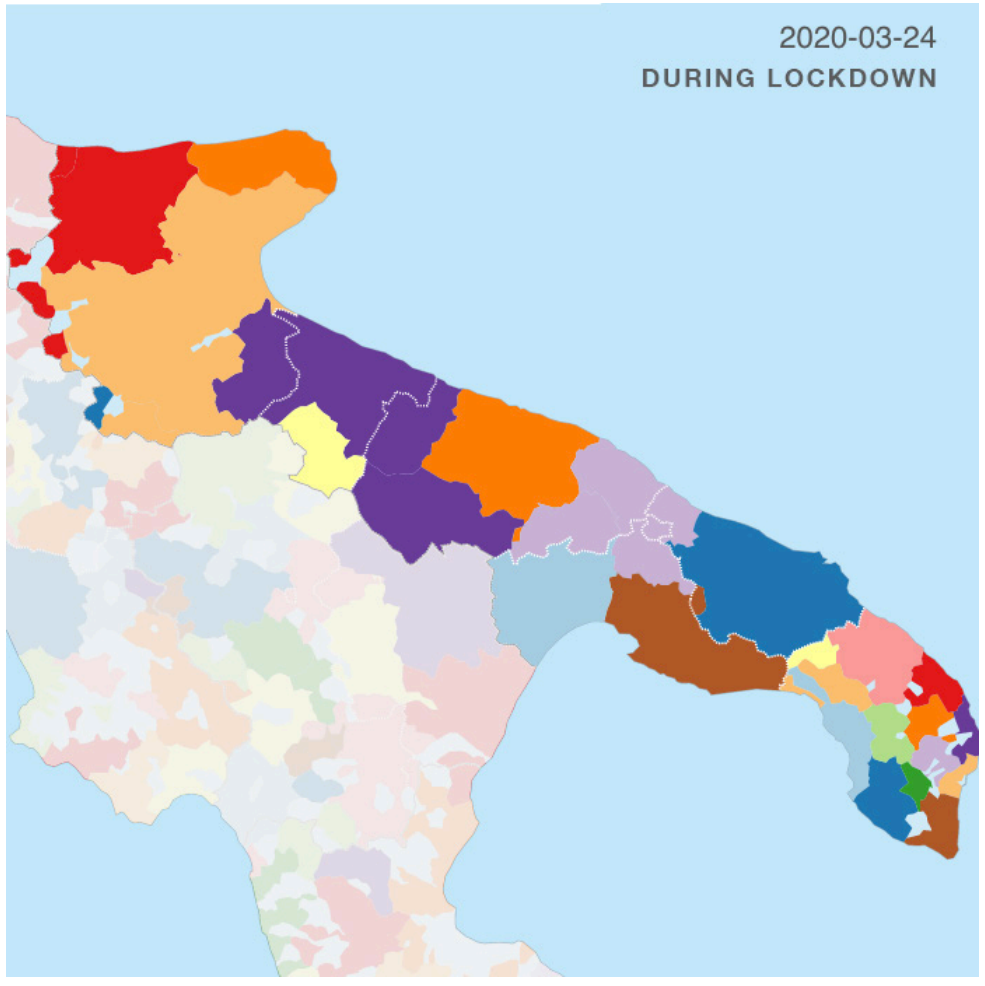

**FIGURE 9 - LOCAL JOB MARKETS IN PUGLIA.**

Local job markets in Puglia (a region in the south-east of the country) before the lockdown (left) and during the lockdown (right).

# CONCLUSIONS

In our first report of the analysis of mobility flows using mobile phone data up to March 28th, 2020, we find several interesting results. First, regarding the volume of in-, out- and self-flows between provinces, we find a significant decrease after the national lockdown (March 9th). Still, we do not find any significant decrease soon after the closure of the non-necessary productive activities.

Regarding the in- and out-flow diversities, we find that while there is a slight increase in the flow diversity on the weekends before the lockdown, there is a strong decrease of the flow diversity on the weekends during the lockdown. Moreover, the application of data mining techniques reveals the presence of five main clusters of provinces.

Finally, we use a community detection algorithm to find local job markets in Italy. We observe a striking increase in the number of communities during the lockdown and a slight increase after the closure of non-necessary activities. This suggests that reduced mobility split the territory into more and smaller local job markets. This information may be exploited by decision- and policy-makers to plan "phase 2" of the management of the epidemics.

In the next report, we will investigate deeply how the structure of the OD matrices evolve in time, and we will extend the period of observation to the most recent days. We will also focus our analysis on some specific regions, considering the evolution of the epidemics at a municipality level. We will compare the impact of mobility reduction to the outbreak, answering several analytical questions: What is the virus-spreading effect generated by late-February north-south flows? How large should a "red zone" be to reduce effectively the spread of the epidemic?

# APPENDIX

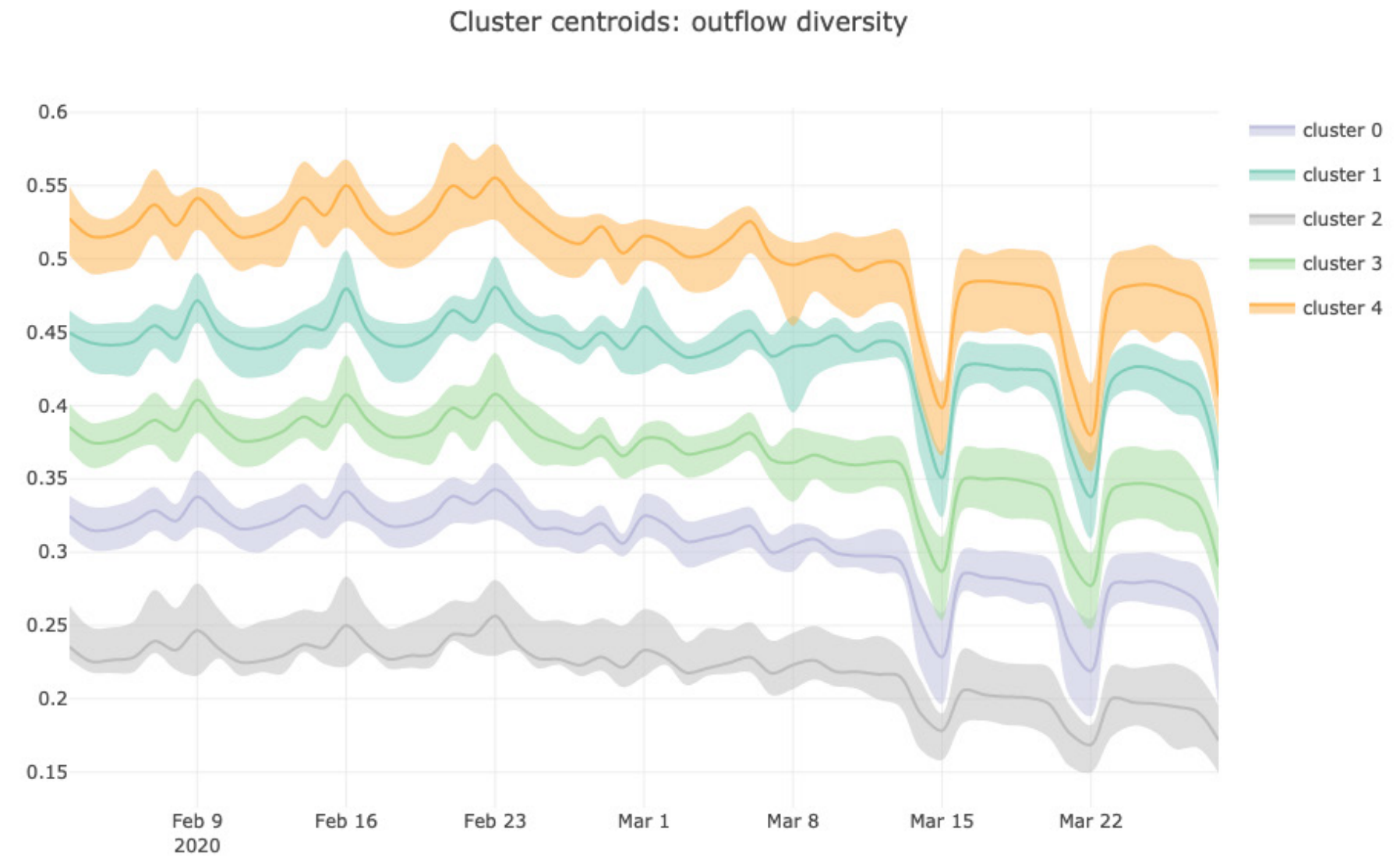

**FIGURE 10**

Evolution of the out-flow diversity of the five clusters' centroids. The area around the line indicates the deviation of the provinces from the centroid. Note that, though the clusters have similar trends, they have different typical out-flow diversities.

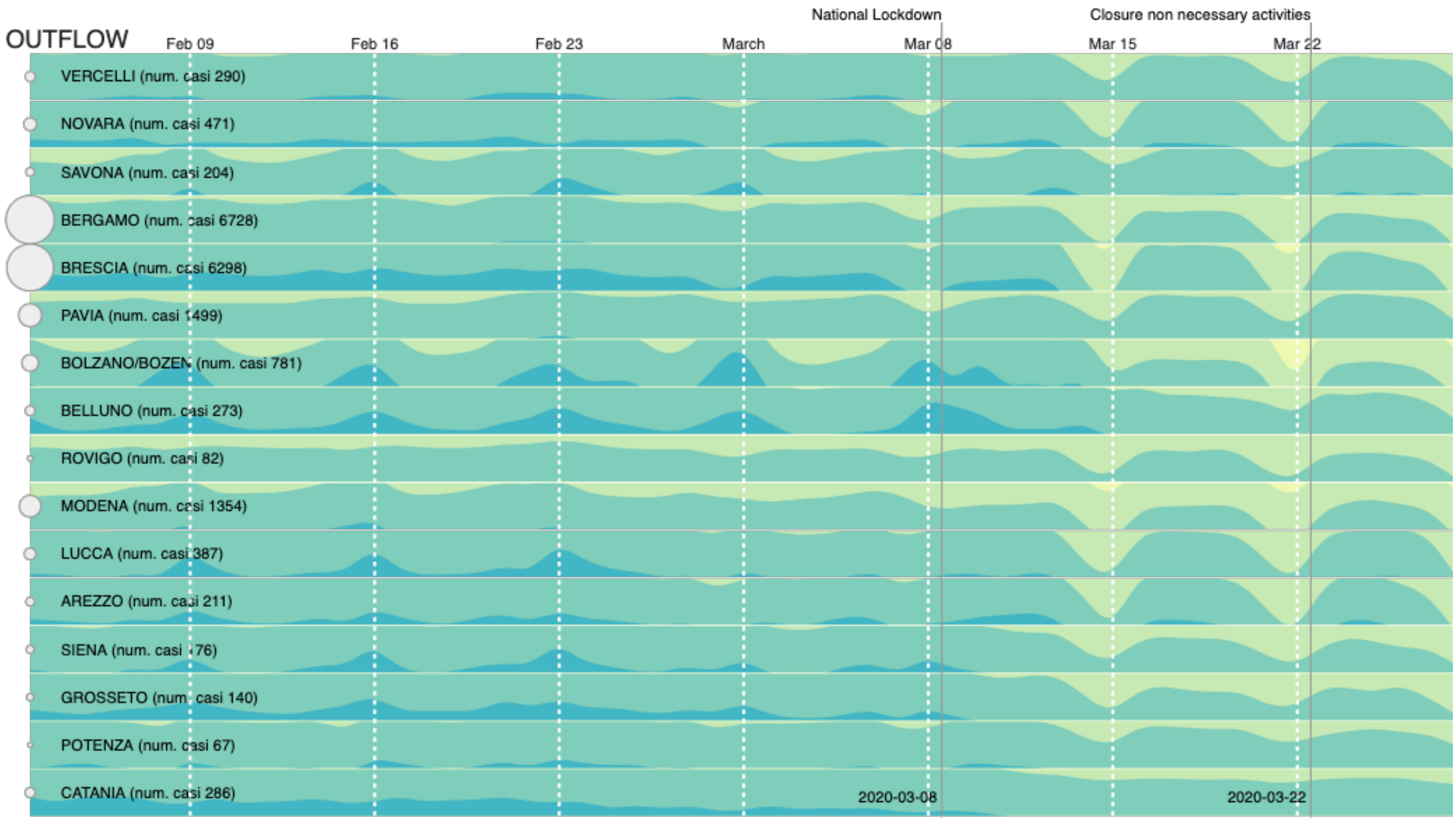

**FIGURE 11**

Figure. Evolution of the in-flow diversity of provinces in cluster 1 This cluster includes Bergamo and Brescia, territories among the most hit by COVID-19. The circles on the left have an area proportional to the number of confirmed cases in the corresponding province up to March 24th.

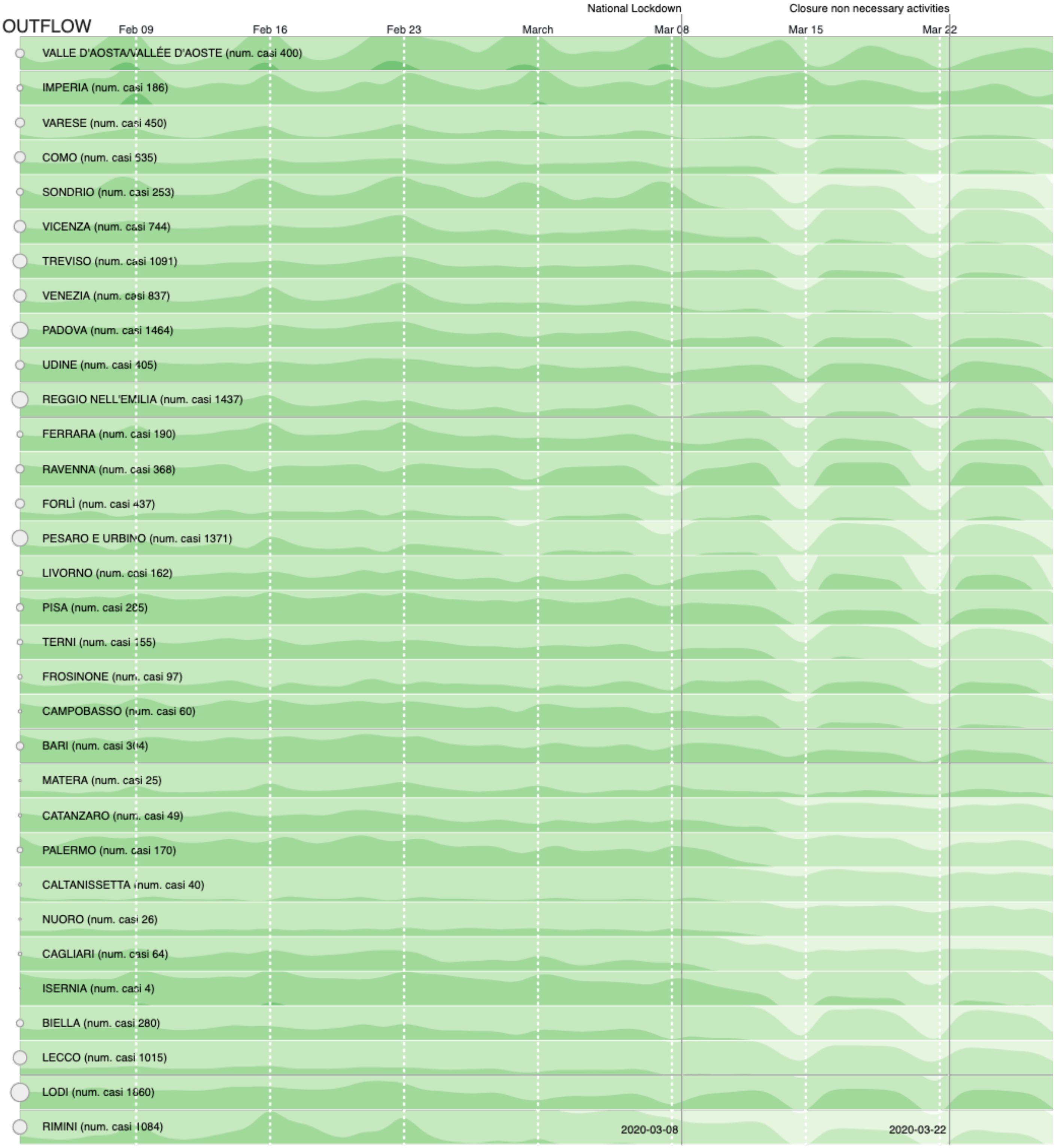

**FIGURE 12**

Evolution of the in-flow diversity of provinces in cluster 3. The circles on the left have an area proportional to the number of confirmed COVID-19 cases in the corresponding province up to March 24th.

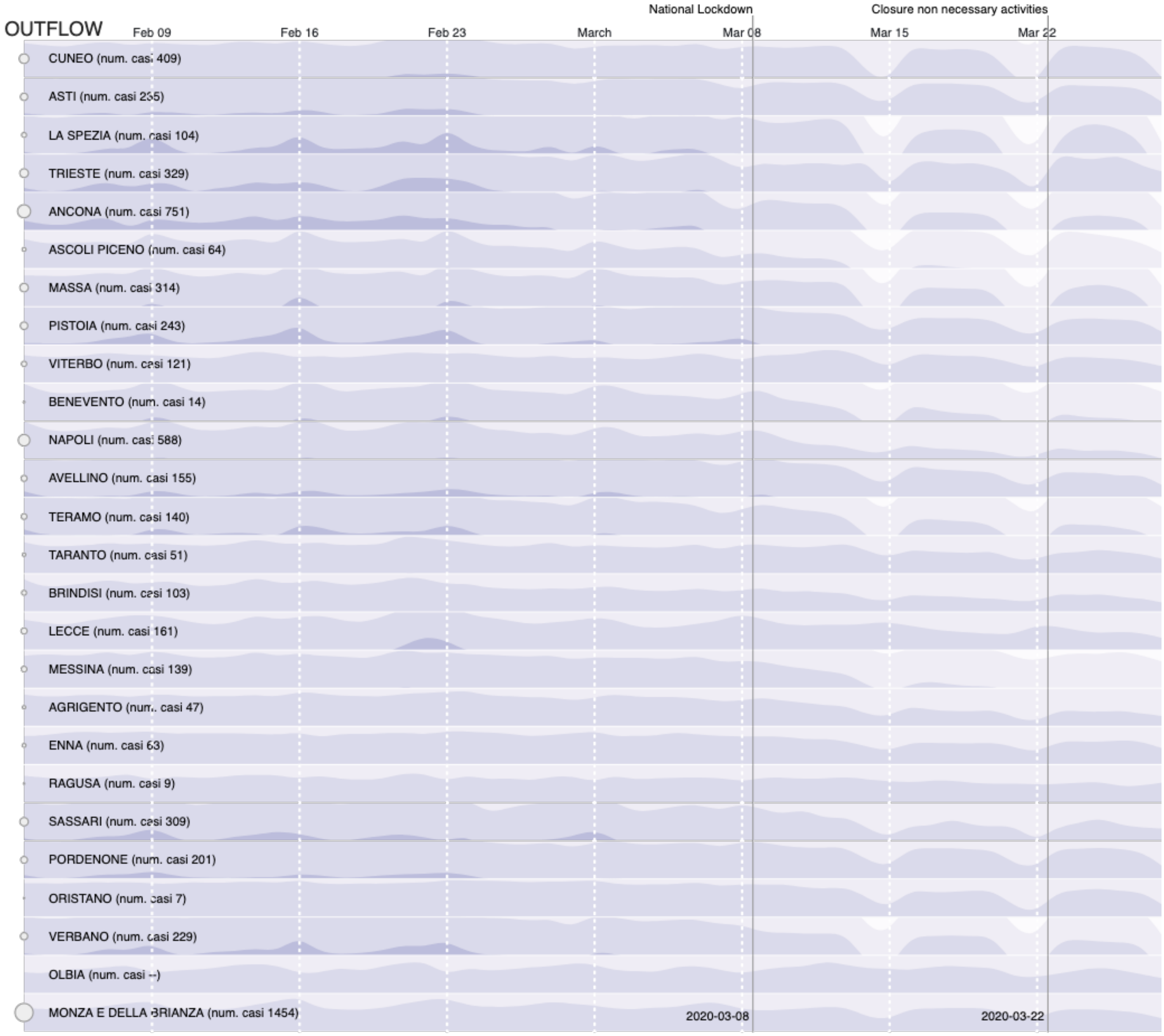

**FIGURE 13**

Evolution of the in-flow diversity of provinces in cluster 0. The circles on the left have an area proportional to the number of confirmed COVID-19 cases in the corresponding province up to March 24th.

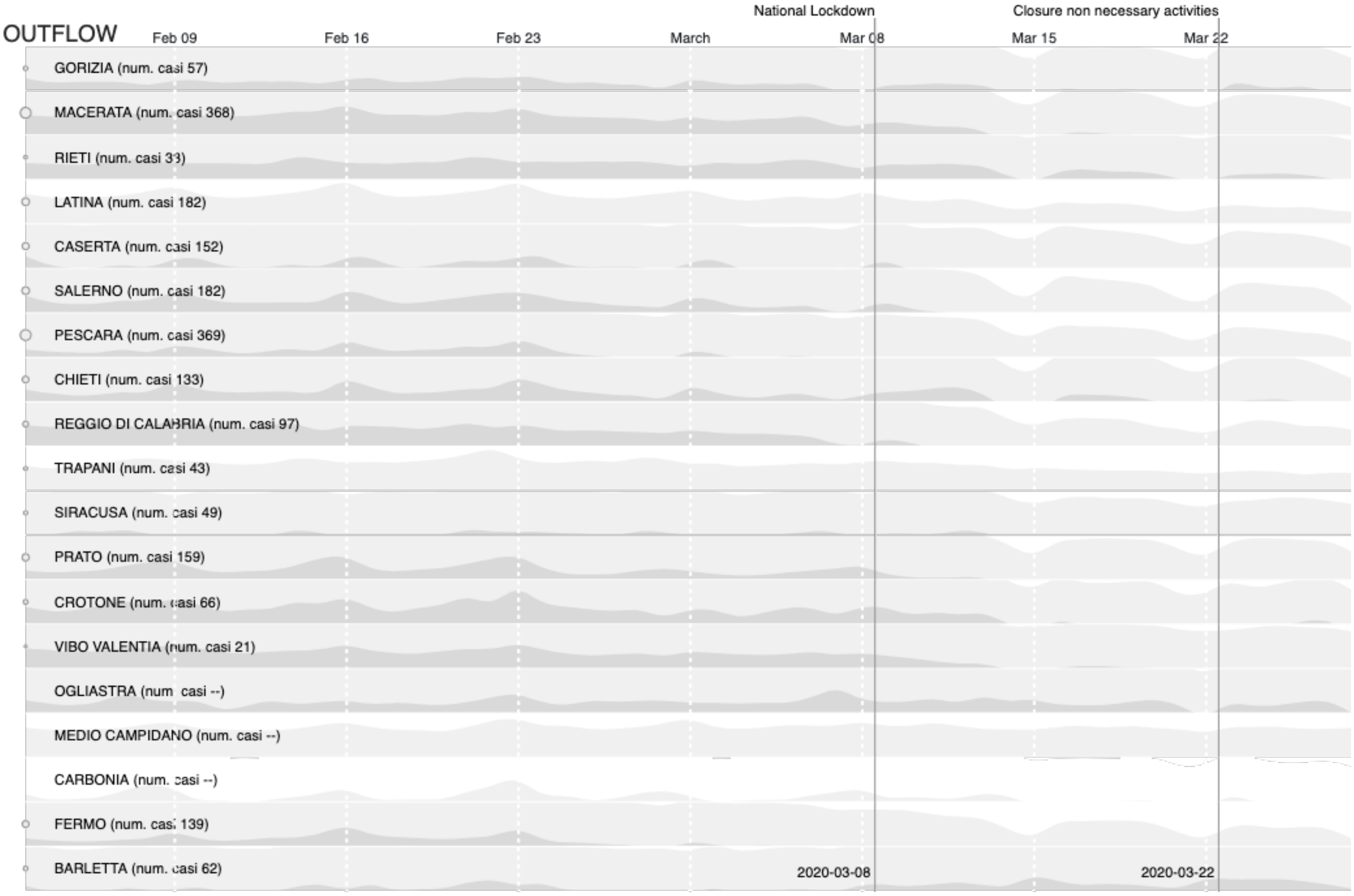

**FIGURE 14**

Evolution of the in-flow diversity of provinces in cluster 2. The circles on the left have an area proportional to the number of confirmed COVID-19 cases in the corresponding province up to March 24th.

Flow diversity and local job markets
during the national lockdown

# MOBILE PHONE DATA ANALYTICS AGAINST THE COVID-19 EPIDEMICS IN ITALY

## ISSUE #1

April 2020

**CNR**

—

ISTI

Fosca Giannotti
Mirco Nanni
Luca Pappalardo
Giulio Rossetti
Salvatore Rinzivillo

**UNIVERSITY OF PISA**

—

COMPUTER SCIENCE DEPARTMENT

Paolo Cintia
Daniele Fadda
Dino Pedreschi

—

DEPARTMENT OF TRANSLATIONAL RESEARCH ON NEW TECHNOLOGIES IN MEDICINE AND SURGERY

Pier Luigi Lopalco
Sara Mazzilli
Lara Tavoschi

**WINDTRE**

—

BIG DATA & ANALYTICS

Pietro Bonato
Francesco Fabbri
Francesco Penone
Marcello Savarese

for more information
rinzivillo@isti.cnr.it, luca.pappalardo@isti.cnr.it,
cintia@di.unipi.it, dfadda@di.unipi.it